\documentstyle[aps,prl,epsf,multicol]{revtex}
\begin{document}
\title{\bf Exact Occupation Time Distribution in a Non-Markovian Sequence and Its
Relation to Spin Glass Models}

\author{Satya N. Majumdar and David S. Dean}
\address{ Laboratoire de Physique Quantique (UMR C5626 du CNRS),
Universit\'e Paul Sabatier, 31062 Toulouse Cedex, France. \\}

\date{3 July 2002}

\maketitle

\begin{abstract} 
We compute exactly the distribution of the occupation time in a discrete {\em non-Markovian}
toy sequence which appears in various physical contexts such as the diffusion processes
and Ising spin glass chains. The non-Markovian property makes the results nontrivial
even for this toy sequence. The distribution is shown to have non-Gaussian tails
characterized by a nontrivial large deviation function which is
computed explicitly. An exact mapping of 
this sequence to an Ising spin glass chain via a 
gauge transformation raises an interesting new question for a generic finite sized
spin glass model: at a given temperature, what is the distribution (over disorder) of the thermally 
averaged
number of spins that are aligned to their local fields? We show that this distribution
remains nontrivial even at infinite temperature and can be computed explicitly
in few cases such as in the Sherrington-Kirkpatrick model with Gaussian disorder.
    
\noindent

\end{abstract}

\begin{multicols}{2}

\section{Introduction}

The occupation time $T$ of a stochastic process $x(t)$ is simply the time 
that the process spends above its mean value (say $0$) when observed
over the period $[0,t]$,
\begin{equation}
T= \int_0^t \theta \left[x(t')\right] dt',
\label{occupdef}
\end{equation}
where $\theta [x]$ is the Heaviside step function and we assume, for simplicity,
that the process starts at $x(0)=0$. Since the seminal work of L\'evy \cite{Levy}
who computed the exact probability distribution of $T$ in the case when $x(t)$
is just an ordinary Brownian motion, there has been a lot of interest in the
mathematics community to study the occupation time for various processes\cite{math,Lamperti}.
Recently the study of the occupation time has seen a revival in the physics 
community in the context of nonequilibrium systems\cite{DGNT,physics} due to its
potential applications in a wide range of physical systems which include, amongst others,
optical imaging\cite{WC}, analysis of the morphology of growing surfaces\cite{TND} and
analysis of the fluorescence intermittency emitting from colloidal semiconductor
dots\cite{ens}.  

The occupation time $T$ is clearly a random variable. Its probability distribution
$P(T,t)$ evidently depends on the window size $t$. It turns out that quite generically there 
are essentially two types of asymptotic behaviors of this distribution $P(T,t)$
depending on whether the underlying stochastic process $x(t)$ is {\em non-stationary}
or {\em stationary}. A non-stationary process is one where the two-time correlation
function $C(t,t')=\langle x(t)x(t')\rangle$ depends on both times $t$ and $t'$.
An example is the ordinary Brownian motion, $dx/dt=\eta(t)$ where $\eta(t)$ is a
Gaussian white noise with $\langle \eta(t)\rangle =0$ and $\langle \eta(t)\eta(t')\rangle 
=\delta(t-t')$. In this case, $C(t,t')={\rm min}(t,t')$. In a stationary process, on
the other hand, the two-time correlation function depends only on the
time difference, $C(t,t')=C(|t-t'|)$. A simple example of a stationary process
is the Orstein-Uhlenbeck process, $dx/dt=-\lambda x +\eta(t)$, where a particle
moves in a parabolic potential in presence of external thermal noise. In this
case, the particle reaches equilibrium at long times when the two-time
correlation simply becomes, $C(t,t')= \exp[-\lambda |t-t'|]$. 

In the non-stationary case, one expects that in the asymptotic limit $t\to \infty$, 
$T\to \infty$ but keeping the ratio $r=T/t$ fixed, the distribution $P(T,t)$ has
the generic scaling behavior,
\begin{equation}
P(T,t) \sim {1\over {t}}f\left({T\over {t}}\right),
\label{scaling1}
\end{equation}
where the scaling function $f(r)$ has nonzero support only in the range $r\in [0,1]$.
For example, in the case of ordinary Brownian motion, the scaling function $f(r)$ 
can be computed exactly\cite{Levy}, $f(r)= 1/{\pi \sqrt{r(1-r)}}$. This is known
as the Arc-Sine law of L\'evy since the cumulative distribution has an arc-sine
form, $\int_0^{r} f(r')dr'=2{\sin}^{-1}\left(\sqrt{r}\right)/{\pi}$. 
Note that for the Brownian case
the scaling actually holds for all $t$ and $T$. The analytical
calculation of this scaling function $f(r)$ is, however, nontrivial
even for this simple Brownian case. Following the work of L\'evy,
there have been various generalizations of this Arc-Sine law. For example,
the scaling function $f(r)$ has been computed exactly for the so called
L\'evy processes\cite{Lamperti}, and recently for a more general class of
renewal processes\cite{GL}. The occupation time 
distribution has also been studied recently for a Brownian particle moving 
in a random Sinai type potential and
the corresponding scaling function $f(r)$ has been computed exactly\cite{MC}.  

For stationary processes, on the other hand, the distribution $P(T,t)$ is
expected to have the following generic asymptotic behavior in the 
appropriate scaling limit $T\to \infty$, $t\to \infty$ with the ratio $r=T/t$
fixed\cite{MB},
\begin{equation}
P(T,t) \sim e^{-t \Phi(T/t)},
\label{scaling2}
\end{equation}
where  $\Phi(r)$ is a large deviation function with, in general, 
non-Gaussian tails\cite{MB}. For example, for the Ornstein-Uhlenbeck stationary
process discussed in the previous paragraph, the function $\Phi(r)$ has
recently been computed exactly by utilizing a mapping to a
quantum mechanical path integral problem\cite{MB}.

The calculation of either the scaling function $f(r)$ for non-stationary
processes or the large deviation function $\Phi(r)$ for stationary processes
is a challenging theoretical problem. So far, exact results exist only
for Markov processes where the value of the process $x(t)$ at time $t$
depends only on its value at the previous time step, say at $t-\Delta t$
where $\Delta t$ is an infinitesimal time step, but is completely  
independent on the previous history of the process. For example, 
the ordinary Brownian motion and the Orstein-Uhlenbeck processes are
both Markovian. On the other hand, most processes in nature are 
non-Markovian and the Markov processes are more of exceptions  
rather than rules. Non-Markov processes
are known to be notoriously difficult for the analytical
calculation of even simpler quantities such as persistence, i.e.,
the probability that the process does not change sign up to 
time $t$\cite{persistence}.
Naturally the analytical calculation of the occupation time
distribution for non-Markovian processes is even more difficult.

For a certain class of `smooth' non-Markovian processes such
as the diffusion equation, it is possible to
compute the occupation time distribution\cite{DGNT} using the 
independent interval approximation (IIA) which assumes
that the intervals between successive zero crossings are statistically  
independent\cite{Diffusion}. But these IIA results are only
approximate. To our knowledge, there exists no exact result
for the occupation time distribution for a non-Markov process,
either stationary or non-stationary. In this paper, we obtain,
for the first time, an exact analytical result for the occupation
time distribution for a stationary non-Markovian process.
To be more precise, we actually study the occupation time
distribution of a discrete stationary non-Markovian sequence
and not a continuous stochastic process. Nevertheless 
the asymptotic behavior as given by Eq. (\ref{scaling2}) still
remains true and we compute analytically the corresponding
large deviation function $\Phi(r)$.

Recently the importance of studying the statistical properties such as
the persistence and the distribution of the number of zeros of 
{\em discrete stochastic sequences},
as opposed to the more traditional {\em continuous stochastic processes}, has
been emphasized in a number of articles\cite{discrete,MD,M}. There are
two principal reasons for studying a stochastic sequence. First, in
various experiments and numerical simulations, even though the underlying
physical process is continuous in time, in practice one actually measures 
the events only at discrete time points. The result of this discretization 
can lead to subtle and important differences between the `true' properties
of the process and the `measured' properties\cite{discrete}. To
estimate these differences, it is important to study the properties
of a discrete sequence. The second reason follows from the observation\cite{MD}
that many processes in nature such as weather records are stationary under
translations in time only by an integer multiple of a basic period. For
example, the seasons repeat typically every one year. For such processes,
it was observed in Ref. \cite{MD} that the
persistence of the underlying continuous process coincides with that of the
discrete sequence obtained from the measurement of the process only
at times that are integer multiples of the basic period.

In particular, in Ref. \cite{MD} a specific discrete sequence was
obtained as a limiting case of the diffusion equation on a hierarchical
lattice. This rather simple {\em toy} sequence, even though non-Markovian, had
the remarkable property of being solvable for certain 
statistical properties such as the persistence\cite{MD} and the
distribution of the number of zeros\cite{M}. Furthermore, these
exact results were rather nontrivial\cite{MD,M} even for this toy sequence. 
It is always important to have
a such a solvable non-Markovian toy model which can then be used as a
benchmark to predict the possible expected behaviors of various observables
in a more complex non-Markovian process. In this paper we show that 
the occupation time distribution can also be computed exactly for this
toy model and like other quantities such as the persistence, it is 
rather nontrivial even for this simple toy model.      

We further make an interesting observation that the occupation time
for this toy sequence is related to a specific physical
observable in an Ising spin glass chain with nearest neighbor 
interactions. In a given sample of
the spin glass chain, one can ask: what is the average (thermal) number
of spins that are aligned to the direction of their local fields? This
physical object is a random variable that fluctuates from one sample
of disorder to another. A natural question is: what is the
probability distribution (over disorder) of this thermal average? It 
turns out that this distribution is nontrivial even at infinite temperature. 
In fact, we show that at infinite temperature this distribution 
in the spin glass chain coincides exactly with the occupation time 
distribution of the toy sequence mentioned above. This connection is useful 
as it raises a general question for any spin glass model (and not just
restricted to a chain): what is the probability distribution of
the average (thermal) number of spins that are aligned to their
local fields? In this paper, we show that the analytical computation
of this distribution in the limit of infinite temperature, though still 
nontrivial, is tractable in few cases. In particular,
we calculate analytically this infinite temperature distribution in
the Sherrington-Kirkpatrick (SK) model of mean field spin glasses\cite{SK}.

The layout of the paper is as follows. In Sec. II, we define the toy sequence,
recall some of its properties and known results and then compute the
occupation time distribution exactly. In Sec. III, we establish the connection
to a spin glass chain and raise the general question regarding the distribution
of the average number of spins aligned to their local fields in a generic
spin glass model. In Sec. IV,
we compute this distribution analytically in the SK model at infinite temperature
and show that it is nontrivial even at infinite temperature. Finally we
conclude in Sec. V with a summary and some open questions.
  
\section{The Toy Sequence and Its Exact Occupation Time Distribution}

The toy sequence we study in this section was originally derived as a limiting case
of the diffusion process on a hierarchical lattice\cite{MD}. This is a sequence $\{\psi_i\}$
of correlated random variables constructed via the following rule,
\begin{equation}
\psi_i=\phi_i+ \phi_{i-1}, \,\,\,\,\,i=$1$,$2$,$\ldots$,$N$,
\label{psi1}
\end{equation}                                                           
where $\phi(i)$'s are independent and identically distributed (i.i.d) random variables, 
each drawn from the same symmetric continuous distribution $\rho(\phi)$. Note that 
even though $\phi(i)$'s are uncorrelated, the variables $\psi_i$'s are correlated.
The two point correlation function, $C_{i,j}=\langle \psi_i \psi_j\rangle$ can be 
easily computed from Eq. (\ref{psi1}),
\begin{equation}
C_{i,j}=\sigma^2\left[2\delta_{i,j} + \delta_{i-1,j} +\delta_{i,j-1}\right],
\label{corr1}
\end{equation}
where $\delta_{i,j}$ is the Kronecker delta function and 
$\sigma^2=\int_{-\infty}^{\infty}\phi^2\rho(\phi)d\phi$ which we assume to be finite.
Thus the sequence $\{\psi_i\}$ has only nearest neighbor correlation. Also note that
for large sequence size $N$, the sequence is {\em stationary} since $C_{i,j}$ depends
only on the difference $|i-j|$, and not individually on $i$ or $j$. The sequence
$\{\psi_i\}$ is also {\em non-Markovian}. To see this, one can try to express 
a specific member of the sequence, say $\psi_i$, only in terms of other members of the 
sequence\cite{MD}. This can be easily done using Eq. (\ref{psi1}) and one gets for any $i\ge 2$,
\begin{equation}
\psi_i = \sum_{k=1}^{i-1} (-1)^{k-1}\psi_{i-k} + \phi_i +(-1)^{i-1}\phi_0 . 
\label{nonmarkov}
\end{equation}
This relation clearly demonstrates the history dependence of the sequence in the sense
that $\psi_i$ depends not just only on the previous member $\psi_{i-1}$ (as would have been
in the Markov case), but on the whole history of the sequence preceding $\psi_i$. 

We now turn to the exact computation of the occupation time distribution for the sequence
in Eq. (\ref{psi1}). The occupation time $R$ in this case is simply the number of $\psi_i$'s
that are positive out of the total number $N$ and is given by the discrete counterpart 
of Eq. (\ref{occupdef}),
\begin{equation}
R = \sum_{i=1}^N \theta \left( \psi_i \right).
\label{occupseq}
\end{equation}  
Clearly $R$ is a random variable over the range $0\le R\le N$. Let us denote its
probability distribution by $P(R,N)$ which is formally given by,
\begin{equation}
P(R,N)=\int \delta\left[ R - \sum_{i=1}^N 
\theta\left(\phi_{i-1}+\phi_i\right)\right]\prod_{i}\rho(\phi_i)d\phi_i.
\label{distri0}
\end{equation}
Analogous to the asymptotic behavior in Eq. (\ref{scaling2}) for continuous stationary processes,
we will show that in the appropriate scaling limit $R\to \infty$, $N\to \infty$ but keeping
the ratio $r=R/N$ fixed, the distribution $P(R,N)$ has the scaling behavior,
\begin{equation}
P(R,N)\sim e^{-N \Phi(R/N)},
\label{scalingseq}
\end{equation}
where the large deviation function $\Phi(r)$ can be computed analytically. Note also
that since $\rho(\phi)$ is symmetric around the origin, the number of 
positive members of the sequence must have the same distribution as the
number of negative members, i.e., $P(R,N)=P(N-R, N)$. Consequently, we must
have $\Phi(r)=\Phi(1-r)$, i.e., the large deviation function over the allowed
range $0\le r\le 1$ must be symmetric around $r=1/2$.

To compute the distribution $P(R,N)$ we use a transfer matrix method which has already
been used successfully to calculate other quantities for this sequence such as the 
persistence\cite{MD} and
the distribution of the number of sign changes\cite{M}. To start with, we define
$Q^{\pm}_{R,N}(\phi_0)$ denoting respectively the joint probability that the first member 
of the sequence $\psi_1$ is positive (negative) and that the sequence of size $N$ has
a total $R$ number of positive members, given the value of $\phi_0$. Let us
also define $Q_{R,N}(\phi_0)= Q^{+}_{R,N}(\phi_0) + Q^{-}_{R,N}(\phi_0)$ which
denotes the probability of having $R$ positive members in a sequence of size $N$,
given $\phi_0$. The required
occupation time distribution is then given by, 
\begin{equation}
P(R,N)= \int_{-\infty}^{\infty} Q_{R,N}(\phi_0)\rho(\phi_0)d\phi_0.
\label{PMN}
\end{equation}
The reason for this small detour is simply that one can write quite easily a recursion relation
for the joint probabilities $Q^{\pm}_{R,N}(\phi_0)$. However it is not easy to write
a recursion directly for the distribution $P(R,N)$. The probabilities
$Q^{\pm}_{R,N}(\phi_0)$ satisfy the following recursion relations,
\begin{eqnarray}
Q^{+}_{R,N}(\phi_0)&=&\int_{-\phi_0}^{\infty}d\phi_1
\rho(\phi_1)Q_{R-1,N-1}(\phi_1) \nonumber \\
Q^{-}_{R,N}(\phi_0)&=&\int_{-\infty}^{-\phi_0}d\phi_1
\rho(\phi_1)Q_{R,N-1}(\phi_1).
\label{recur1}
\end{eqnarray}     
The above recursion relations are valid for all $0\le R \le N$ and $N\ge 1$ with the 
initial conditions $Q^{+}_{0,0}(\phi_0)=1$ and $Q^{-}_{0,0}(\phi_0)=0$. 

These recursion relations in Eq. 
(\ref{recur1}) are easy to follow. Consider first the relation for $Q^{+}_{R,N}(\phi_0)$.
In order for the first member $\psi_1$ to be positive, it follows from the definition,
$\psi_1=\phi_1+\phi_0$, that $\phi_1>-\phi_0$ for a given $\phi_0$. This explains
the integration range on the right hand side of the first line in Eq. (\ref{recur1}).
Also once the first member is positive, in order to have a total $R$ positive members,
we need to ensure that the rest of the chain of size $N-1$ (excluding the first member)
has exactly $R-1$ positive members. The probability of this latter event, for a given
$\phi_1$, is simply $Q_{R,N}(\phi_1)$. This explains the integrand on the right hand
side of Eq. (\ref{recur1}). The second line of Eq. (\ref{recur1}) can be understood 
following a similar line of reasoning. Note that the recursion relations in Eq. (\ref{recur1})
also satisfy the one sided boundary conditions, $Q^{+}_{R,N}(-\infty)=0$ and
$Q^{-}_{R,N}(\infty)=0$. The first condition follows from the fact that if $\phi_0\to -\infty$,
then the first member of the sequence $\psi_1=\phi_1+\phi_0$ can be positive only
with a vanishing probability. On the other hand if $\phi_0\to \infty$, then $\psi_1$
can be negative only with probability zero thus giving rise to the second condition. 
Note however that the values at the other boundaries namely $Q^{+}_{R,N}(\infty)$
and $Q^{-}_{R,N}(-\infty)$ are unspecified.

We next define the generating functions, 
\begin{equation}
{\tilde Q}^{\pm}_{N}(\phi_0,y)=\sum_{R=0}^{\infty} Q^{\pm}_{R,N}(\phi_0)y^R,
\label{genfunc}
\end{equation}
with the understanding that $Q^{\pm}_{R,N}(\phi_0)=0$ for $R>N$ since $R$ can take
values only in the range $0\le R\le N$. We also define ${\tilde Q}_{N}(\phi_0,y)=
{\tilde Q}^{+}_{N}(\phi_0,y)+{\tilde Q}^{-}_{N}(\phi_0,y)$. Using Eq. (\ref{recur1}), it is
easy to see that the generating functions satisfy the recursions,
\begin{eqnarray}
{\tilde Q}^{+}_{N}(\phi_0,y)&=&y\int_{-\phi_0}^{\infty}d\phi_1
\rho(\phi_1){\tilde Q}_{N-1}(\phi_1,y) \nonumber \\
{\tilde Q}^{-}_{N}(\phi_0,y)&=&\int_{-\infty}^{-\phi_0}d\phi_1
\rho(\phi_1){\tilde Q}_{N-1}(\phi_1,y),
\label{recur2}
\end{eqnarray} 
with the boundary conditions ${\tilde Q}^{+}_N(-\infty,y)=0$ and
${\tilde Q}^{-}_{N}(\infty,y)=0$ for all $y\ge 0$. These generating functions
also satisfy the condition ${\tilde Q}_0(\phi_0,y)=1$ for all $y$. The next step
is to differentiate the recursion relations in Eq. (\ref{recur2}) with respect to
$\phi_0$ which gives,
\begin{eqnarray}
{{\partial {\tilde Q}^{+}_{N}(\phi_0,y)}\over {\partial \phi_0}}&=&y
\rho(-\phi_0){\tilde Q}_{N-1}(-\phi_0,y) \nonumber \\
{{\partial {\tilde Q}^{-}_{N}(\phi_0,y)}\over {\partial \phi_0}}
&=&- \rho(-\phi_0){\tilde Q}_{N-1}(-\phi_0,y).
\label{recur3}
\end{eqnarray} 
Further simplifications can be made by using the symmetry $\rho(-\phi_0)=\rho(\phi_0)$
and by making a change of variable from $\phi_0$ to $u(\phi_0) =\int_0^{\phi_0}\rho(\phi)d\phi$.
Note that since $\rho(\phi)$ is symmetric around the origin, $\phi_0 \to -\phi_0$ corresponds to
$u \to -u$. Thus $u(\phi_0)$ is a monotonic function of $\phi_0$. 
Note further that as $\phi_0\to \pm\infty$, $u\to \pm 1/2$, where we have again used
the fact that $\rho(\phi_0)$ is symmetric around the origin.
Let us also
write, ${\tilde Q}^{\pm}_{N}(\phi_0,y) =S^{\pm}_{N}(u,y)$ and 
${\tilde Q}_{N}(\phi_0,y) =S_{N}(u,y)$ where $S_{N}(u,y)=S^{+}_{N}(u,y)+S^{-}_{N}(u,y)$.
Then the relations in Eq. (\ref{recur3}) simplify to,
\begin{eqnarray}
{{\partial {S}^{+}_{N}(u,y)}\over {\partial u}}&=&y S_{N-1}(-u,y) \nonumber \\
{{\partial {S}^{-}_{N}(u,y)}\over {\partial u}}
&=&-S_{N-1}(-u,y), 
\label{recur4}
\end{eqnarray}  
which are valid over $-1/2\le u \le 1/2$. In terms of the variable $u$, the boundary conditions
${\tilde Q}^{+}_N(-\infty,y)=0$ and
${\tilde Q}^{-}_{N}(\infty,y)=0$ translate to
$S^{+}_N(-1/2,y)=0$ and $S^{-}_N(1/2,y)=0$ for all $y\ge 0$.
Note also the interesting fact that the distribution $\rho(\phi)$ has completely disappeared
in Eq. (\ref{recur4}). The consequence of this, as we will see later, is that occupation
time distribution $P(R,N)$ is completely universal, i.e., independent of the
distribution $\rho(\phi)$ as long as it is symmetric and continuous.   

The recursion relations in Eq. (\ref{recur4}), though much simplified, are still nontrivial 
since they are nonlocal in $u$. We next employ the technique of separation of variables,
$S^{\pm}_N(u,y)={\lambda}^{-N}f^{\pm}(u)$
where we have suppressed the $y$ dependence for convenience of notations. Substituting 
this form in Eq. (\ref{recur4}), we get a non-local eigenvalue equation,
\begin{eqnarray}
{ {d f^{+}}\over {du}}&=& y\lambda \left[f^+(-u)+f^{-}(-u)\right] \nonumber \\
{ {d f^{-}}\over {du}}&=& -\lambda \left[f^+(-u)+f^{-}(-u)\right],
\label{eigen1}
\end{eqnarray} 
where the eigenvalue $\lambda$ is yet to be determined. We also have the boundary conditions,
$f^+(-1/2)=0$ and $f^{-}(1/2)=0$. It is easy to see from Eq. (\ref{eigen1}) that the
sum, $f(u)=f^+(u)+f^{-}(u)$ satisfies the non-local first order equation, $f'(u)=-\omega f(-u)$
where $\omega=\lambda(1-y)$. Differentiating this equation once more, we get a
local second order equation, $f''(u)=-\omega^2 f(u)$ whose most general solution
is given by $f(u)= A \cos (\omega u) + B \sin(\omega u)$ where $A$, $B$ are arbitrary
constants. One further notices that this general solution will also satisfy the first order 
non-local equation $f'(u)=-\omega f(-u)$
provided $B=-A$. Thus we arrive at the solution, 
$f(u)=A\left[ \cos (\omega u)-\sin (\omega u)\right]$.
Substituting this solution on the right hand side of the first line in Eq. (\ref{eigen1})
and solving the resulting equation using the boundary condition $f^{+}(-1/2)=0$, we 
get
\begin{equation}
f^{+}(u)= {{A\lambda y}\over {\omega}}\left[ \sin(\omega u)-\cos (\omega u) +\sin(\omega/2)+\cos 
(\omega/2)\right].
\label{fplus}
\end{equation}   
The other function $f^{-}(u)$ then follows from the relation, $f^{-}(u)=f(u)-f^{+}(u)$
where $f(u)=A[\cos (\omega u)-\sin(\omega u)]$ and $f^{+}(u)$ is given by Eq. (\ref{fplus}).
The function $f^{-}(u)$ still has to satisfy the boundary condition $f^{-}(1/2)=0$.
In fact, this condition determines the eigenvalue $\lambda$ and we get
$\tan (\omega/2) = (1-y)/(1+y)$ where $\omega=\lambda(1-y)$. For large $N$, only the
smallest eigenvalue $\lambda$ will dominate which is given by
\begin{equation}
\lambda = {2\over {(1-y)}}{\tan}^{-1}\left({{1-y}\over {1+y}}\right).
\label{eigenv}
\end{equation}

Using the exact $\lambda(y)$ from Eq. (\ref{eigenv}), we are now
ready to compute the large $N$ behavior of the occupation time distribution $P(R,N)$.
In Eq. (\ref{PMN}), after making a change of variable $\phi_0\to u$, we
find the generating function, $\sum_{M} P(R,N)y^R = \int_{-1/2}^{1/2} S_N(u,y)du$.
We substitute the large $N$ behavior $S_N(u,y)\approx {\lambda}^{-N}f(u)$ and
carry out the integration using the exact expression of $f(u)$ to obtain
the following exact large $N$ result,
\begin{equation}
\sum_{M=0}^{\infty} P(R,N)y^R \approx {{2A}\over 
{\omega}}{\sin(\omega/2)}{\left[\lambda(y)\right]}^{-N},
\label{asymp1}
\end{equation}
where $\lambda(y)$ is given by Eq. (\ref{eigenv}). By inverting the generating function 
and carrying out a standard steepest decent analysis for large $N$, large $R$ but keeping
the ratio $r=R/N$ fixed, we get the 
desired result, $P(R,N)\sim \exp\left[ -N \Phi(R/N)\right]$ where the large
deviation function $\Phi(r)$ is given by the exact formula
\begin{equation}
\Phi(r) = \max_y \left[\log \left( {{2y^r}\over {(1-y)}}{\tan}^{-1}\left({{1-y}\over 
{1+y}}\right)\right) \right].
\label{ldf} 
\end{equation}

We first note that the function $Y(y,r)= 2y^r {\tan}^{-1}\left[(1-y)/(1+y)\right]/(1-y)$
inside the `log' in Eq. (\ref{ldf}) is invariant under the transformation
$y\to 1/y$ and $r\to 1-r$, i.e., $Y(y,r)=Y(1/y,1-r)$. This obviously indicates
that $\Phi(r)=\Phi(1-r)$ as expected. Determining $\Phi(r)$ in closed form
seems difficult, though it can be obtained quite trivially using Mathematica,
as displayed by the solid line in Fig. 1.
\begin{figure}
\narrowtext\centerline{\epsfxsize\columnwidth \epsfbox{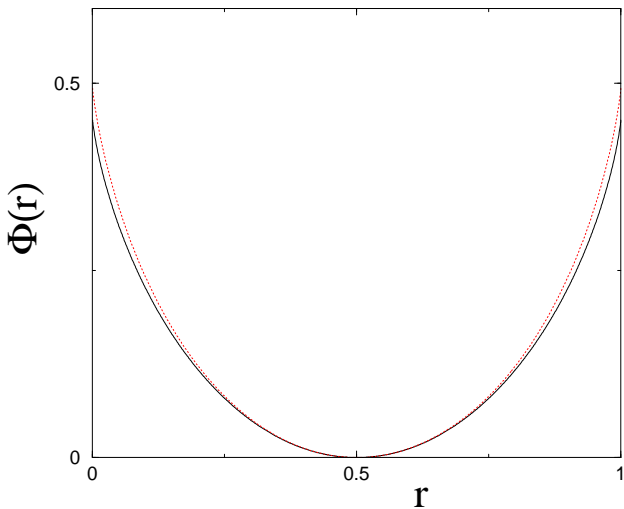}}
\caption{The large deviation function $\Phi(r)$ plotted against $r$.
The solid line corresponds to the large deviation function for the $1$-d sequence 
and is obtained
from Eq. (\ref{ldf}) using Mathematica. The dotted line corresponds to that of the
SK model obtained using Mathematica in Eq. (\ref{thetam}) after the shift $\Phi(r)=\Theta(2r-1)$.}
\end{figure}  

It is however easy and instructive to obtain analytical expressions
of $\Phi(r)$ in the regimes near $r=0$ and $r=1/2$. It turns out
that these limits correspond respectively to $y\to 0$ and $y\to 1$ 
in the function $Y(y,r)$. 
Keeping $r$
fixed we expand $Y(y,r)$ for small $y$ and near $y\to 1$ in a Taylor series and then take the
logarithm and maximize to obtain the following limiting behaviors, 
\begin{eqnarray}
\Phi(r) &=&  \log \left( {\pi \over {2}}\right) + r\log \left( 
{{r\pi}\over {(4-\pi)e}}\right)+ \dots , r\to 0 \nonumber \\
&=& {6\over {5}}{\left( r- {1\over {2}}\right)}^2 + \dots, r\to 1/2.
\label{limiting1}
\end{eqnarray} 
These limiting forms have interesting physical implications. Consider first
the limit $r\to 0$ or equivalently $R\to 0$. Note that 
$P(0,N)=P(N,N)\sim \exp[-\Phi(0)N]$ is just the
probability that all the members are either negative or positive up to length $N$
which is precisely the persistence of the sequence. The persistence
for this sequence was earlier computed in Ref. \cite{MD} and it was found to
decay for large $N$ as $\exp[-\theta N]$ with the persistence exponent
$\theta=\log(\pi/2)$. Thus the limiting form of $\Phi(r)$ as $r\to 0$
in Eq. (\ref{limiting1}) is consistent with the persistence exponent,
$\theta=\Phi(0)=\log(\pi/2)$.   

The other limit $r\to 1/2$ is also interesting and can be derived 
independently from a central limit theorem. To see this
we find from Eq. (\ref{occupseq}) that $R-\langle R\rangle =\sum_{i=1}^N (x_i -\langle 
x_i\rangle)$
where $\langle R\rangle =N/2$, $x_i=\theta(\psi_i)$ and $\langle x_i\rangle =1/2$.
In general the summands $(x_i-\langle x_i\rangle)$ are of course highly correlated
and one can not employ the central limit theorem to evaluate the sum. However, one
can do so in the limit when $M\to \langle M \rangle$ when the variables
$(x_i-\langle x_i\rangle)$ become only weakly correlated. Then the central limit
theorem predicts a Gaussian distribution for the sum, $P(R,N) \sim 
\exp\left[-(R-N/2)^2/2\sigma_N^2\right]$ where $\sigma_N^2 = \langle (R-N/2)^2\rangle$
is the variance. One can calculate this variance independently by computing the
correlation functions $\langle (x_i-1/2)(x_j-1/2)\rangle$ where $x_i=\theta(\psi_i)$.
It is shown in the Appendix that for large $N$, $\sigma_N^2 = 5N/12$. Hence
the central limit theorem predicts that in the limit $R\to 1/2$,
$P(R,N)\sim \exp\left[-6N(r-1/2)^2/5\right]$ thus yielding
exactly the same limiting form of $\Phi(r)$ for $r\to 1/2$ as in
Eq. (\ref{limiting1}).

Thus the occupation time distribution $P(R,N)$, though Gaussian near the mean value $R=N/2$,
becomes non-Gaussian as $r=R/N$ deviates away from its mean and approaches the tails
$r\to 0$ or $r\to 1$.  This crossover from Gaussian behavior near $r=1/2$ to
non-Gaussian behavior near $r\to 0,1$ is characterized by the large deviation function
$\Phi(r)$ changing from a quadratic function near $r=1/2$ to non-quadratic behavior
near $r\to 0,1$ as given by Eq. (\ref{limiting1}). We conclude this section by noting 
the important fact that the large deviation function $\Phi(r)$ in Eq. (\ref{ldf}) and in fact the 
full occupation 
time distribution $P(R,N)$ is completely universal, i.e., independent of
the distribution $\rho(\phi)$ (as long as $\rho(\phi)$ is symmetric and continuous).
Moreover, this universality holds for any arbitrary $N$ and not just asymptotically 
for large $N$.

\section{Relation to Spin Glass Models}

We start this section by raising a physical question for a general
spin glass model defined on a finite lattice of $N$ sites: 
What is the distribution (over disorder) of
the thermally averaged number of spins that are aligned to their
local fields? This distribution depends on the temperature and on
the system size $N$. It turns
out that the distribution remains nontrivial even in the infinite
temperature limit. In fact, for a nearest neighbor Ising spin 
glass chain, we show that this infinite temperature limiting
distribution is precisely that of the occupation time distribution 
of the toy sequence computed in the previous section. The infinite temperature limit, though
nontrivial, is tractable in few other cases such as the SK model
of Ising spin glass which will be discussed in detail in the next section.

Consider a spin glass model on a lattice of $N$ sites defined by the Hamiltonian,
\begin{equation}
E= - \sum_{<i,j>} J_{i,j}S_i S_j,
\label{hamil}
\end{equation} 
where $S_i$'s are the spin variables (not necessarily Ising) and $J_{i,j}$
denotes the coupling between site $i$ and site $j$. In the nearest neighbor
model, the sum in Eq. (\ref{hamil}) runs over nearest neighbor pairs. On the
other hand, for long range mean field models such as the SK model, the sum runs over all
pairs of sites. The variables $J_{i,j}$'s are independent and each is
drawn from the identical distribution $\rho(J)$, which we assume to be
symmetric and continuous. Henceforth we will use the short hand 
notation $\vec J$ and $\vec S$ to denote respectively the set
of couplings and the set of spins. Thus the $J$'s have the joint distribution
$Q\left[\vec J\right]d\vec J = \prod_{i,j}\rho(J_{i,j})dJ_{i,j}$. The local field 
that a spin at site $i$ sees
is simply $h_i= \sum_j J_{i,j}S_j$. If the spin gets aligned to its
local field, we must have $h_iS_i >0$. Hence the total number of spins
$N_a\left[\vec J, \vec S\right]$ in a given configuration that are aligned to their local 
fields
can be formally written as,
\begin{eqnarray}
N_a\left[\vec J, \vec S\right]&=& \sum_i \theta [h_iS_i] \nonumber \\ 
&=&\sum_{i} \theta \left[ S_i\sum_j J_{i,j}S_j\right].
\label{align1}
\end{eqnarray}
Evidently $N_a$ is a random variable that depends on the couplings $\vec J$ as
well as the spins $\vec S$. 
Let us first compute the thermal average of $N_a$ over the spin configurations
for a fixed quenched disorder $\vec J$,
\begin{equation}
{\overline N_a} \left(\vec J\right)= {1\over {Z}} \sum_{\vec S}N_a\left[\vec J,\vec 
S\right]e^{-\beta 
E(\vec S)},
\label{thermal1}
\end{equation}
where $Z=\sum_{\vec S}e^{-\beta E(\vec S)}$ is the partition function and $\beta$ is the
inverse temperature. This thermal average ${\overline N_a}\left(\vec J\right)$ is 
a random variable that 
varies from one realization of disorder to another. We then ask: what is the probability
distribution of this random variable (over disorder) at a given inverse
temperature $\beta$? This probability distribution ${\rm Prob}({\overline N_a} 
=R)=P_{\beta}(R,N)$ can be formally represented as,
\begin{equation}
P_{\beta}(R,N) = \int \delta\left[R - {\overline N_a} 
\left(\vec J\right)\right] Q\left[\vec J\right]d{\vec J}.
\label{distri1}
\end{equation}

The analytical calculation of $P_{\beta}(R,N)$ at arbitrary $\beta$ seems difficult.
Let us, therefore, consider a simpler limit, namely the limit of infinite temperature or 
equivalently $\beta \to 0$. In this limit, the thermal average in Eq. (\ref{thermal1})
becomes simple, ${\overline N_a} \left(\vec J\right)= \sum_{\vec S} N_a[\vec J, \vec S]/N_C$
where $N_C$ is the total number of spin configurations. Thus all spin configurations
are equally likely.
However, the distribution $P_0(R,N)$ as in Eq. (\ref{distri1}), 
even in this infinite temperature limit, is still nontrivial. 

Let us now focus on
Ising spins where $S_i=\pm 1$. Here $N_C=2^N$ where $N$ is total number of lattice
sites. The Eq. (\ref{distri1}), using Eq. (\ref{align1}), then becomes simpler 
for the Ising case, $P_0(R,N)= \int \delta\left[R - {1\over {2^N}}\sum_{\vec S}\sum_{i=1}^N \theta
\left(\sum_{j} J_{i,j}S_iS_j\right)\right]Q\left[\vec J\right]d \vec J $. 
The next step is to make a 
gauge transformation, $\phi_{i,j}= J_{i,j}S_iS_j$. Since the spins are
Ising, i.e., $S_i=\pm 1$, $\phi_{i,j}$'s have the same distribution
as the $J_{i,j}$'s. The advantage of this gauge transformation is that one can then
do away with the configuration sum over the spins and we simply get,
\begin{equation}  
P_0(R,N)=\int \delta\left[ R - \sum_{i=1}^N \theta \left(\sum_{j\ne i} 
\phi_{i,j}\right)\right]Q\left[\vec 
\phi\right] d{\vec \phi}.
\label{gauge}
\end{equation} 

Now consider the special case of a nearest neighbor Ising spin glass chain of size $N$
where $E=-\sum_{i}J_{i,i+1}S_iS_{i+1}$ with free boundary conditions. Various 
properties of this spin glass chain such as the statistics of
the number of metastable states have been studied analytically by Li\cite{Li}
and by Derrida and Gardner\cite{DG}. Recently it was also shown that the
persistence in the toy sequence studied in this paper is the same
as the average fraction of metastable spins in the Ising chain\cite{MD}. 
In the present context, we find the distribution $P_0(R,N)$ in Eq. (\ref{gauge})
reduces to,
\begin{equation}
P_0(R,N)= \int \delta\left[R-\sum_{i=1}^N 
\theta\left(\phi_{i-1}+\phi_{i}\right)\right]Q\left[\vec 
\phi\right]d\vec \phi.      
\label{1dchain}
\end{equation}
Comparing Eqs. (\ref{1dchain}) and  (\ref{distri0}) one immediately finds
$P_0(R,N)=P(R,N)$, where $P(R,N)$ is precisely the occupation time
distribution that was computed in Sec. II. 
This thus establishes the promised link between the
spin glass problem discussed in this section and the non-Markovian
toy sequence discussed in Sec. II. The infinite temperature distribution
(over disorder) of the thermally averaged number of locally aligned spins 
is identical to that of the occupation time distribution of the toy
sequence discussed in Sec. II. From the exact results of $P(R,N)$ derived
in Sec. II, one therefore knows the distribution $P_0(R,N)$ exactly as well.

A question naturally arises: Are there other solvable cases for $P_0(R,N)$
apart from the $1$-d chain? In the next section we show that indeed the
infinite range SK model is one such case where  
the distribution $P_0(R,N)$ can be computed analytically.

\section{The SK model}

In this section we calculate the infinite temperature distribution $P_0(R,N)$
of the thermally averaged number of locally aligned spins in the infinite
range SK model
defined by the Hamiltonian in Eq. (\ref{hamil}) where $\langle i,j\rangle$ 
runs over all pairs of the total number of $N$ sites. The couplings
$J_{i,j}$'s are independent of each other and we assume that each is drawn 
from a Gaussian distribution, $\rho(J)= \sqrt{N/{2\pi}}e^{-NJ^2/2}$. The choice
$J\sim N^{-1/2}$ is necessary to ensure that the free energy is extensive in
the large $N$ limit. It is clear from Eq. (\ref{gauge}) that if we define
$\psi_i=\sum_{j\ne i}\phi_{i,j}$ where each of the $\phi_{i,j}$'s are
independent Gaussian variables with the distribution $\rho(\phi)=\sqrt{N/{2\pi}}e^{-N\phi^2/2}$,
then $R= \sum_{i=1}^N \theta(\psi_i)$. It turns out that for technical reasons it is
easier to consider the variable 
$M=\sum_{i=1}^N {\rm sgn} (\psi_i)$ where ${\rm sgn}(x)=2\theta (x)-1$.
Hence $M=2R-N$. In what follows we will first compute the distribution $P_0(M,N)$
and derive the corresponding distribution of $R$ using the simple shift $M=2R-N$.

Since we are eventually interested in the limit $N\to \infty$, $M\to \infty$ but keeping
the ratio $m=M/N$ fixed, we set $M=mN$ and write
\begin{eqnarray}
P_0(m,N)&=& {\left \langle \delta\left[\sum_{i=1}^N 
{\rm sgn}(\psi_i)-mN\right] \right \rangle}_{\psi} 
\nonumber 
\\
&=&\int_{-\infty}^{\infty} {{d\mu}\over {2\pi}}e^{-i\mu mN} {\left\langle e^{i\mu\sum_{i=1}^N 
{\rm sgn}(\psi_i)}\right \rangle}_{\psi},
\label{deltarep}
\end{eqnarray}  
where we have used the representation of the delta function, $\delta (x) =\int_{-\infty}^{\infty}
e^{i\mu x}d\mu/{2\pi}$ and ${\left\langle \right\rangle}_{\psi}$ denotes the expectation over
the distributions of $\psi_i$'s. Using the identity, ${\left \langle e^{i\mu\, {\rm 
sgn}(y)}\right\rangle}_y = \sum_{\sigma =-1,1}e^{i\mu \sigma} {\left \langle 
\theta(y\sigma)\right\rangle}_y$, one can rewrite Eq. (\ref{deltarep}) as
\begin{equation}
P_0(m,N)=\int_{-\infty}^{\infty} {{d\mu}\over {2\pi}}e^{-i\mu m N}\sum_{ \{\sigma_i=-1,1\}}
{\left\langle \prod_{i}e^{i\mu \sigma_i}\theta(\psi_i\sigma_i)\right\rangle}_{\psi}.
\label{sigmarep}
\end{equation}
We next use the representation, $\theta(x) =\int_0^{\infty} dl \int_{-\infty}^{\infty}{{d\lambda}\over 
{2\pi}}e^{i\lambda (x-l)}$ in Eq. (\ref{sigmarep}), make the transformation $\lambda_i 
\sigma_i\to \lambda_i$ and $\mu\to -\mu$ and then sum over the $\sigma_i$ variables to obtain
\begin{eqnarray}
\lefteqn{ P_0(m,N)=\int_{-\infty}^{\infty} {{d\mu}\over {2\pi}} e^{i\mu m N}\mbox{$\times$} 
}\quad\nonumber \\
&&\mbox{$\times$} \prod_{i} 2\cos (\mu+\lambda_i l_i)
{\left\langle \left[ \int_0^{\infty} dl_i \int_{-\infty}^{\infty} {{d\lambda_i}\over {2\pi}} 
e^{i\lambda_i \psi_i} \right] \right\rangle }_{\psi} .
\label{cosrep}
\end{eqnarray} 

The next step is to evaluate the the expectation value ${\left \langle \prod_i e^{i\lambda_i 
\psi_i}\right\rangle}_{\psi}$. Using $\psi_i = \sum_{j\ne i}\phi_{i,j}$, we note that
$\prod_i e^{i\lambda_i \psi_i} =\prod_{i<j}e^{i(\lambda_i+\lambda_j)\phi_{i,j}}$. 
Using the Gaussian distribution $\rho(\phi_{i,j})$, one can easily evaluate
the expectation value to finally obtain, ${\left \langle \prod_i e^{i\lambda_i
\psi_i}\right\rangle}_{\psi}= \exp\left[-\sum_{i,j}(\lambda_i+\lambda_j)^2/{4N}\right]$.
We next expand the sum, $\sum_{ij}(\lambda_i+\lambda_j)^2 = 2N\sum_i\lambda_i^2 + 2(\sum_i \lambda_i)^2$
and use a Hubbard-Stratonovich transformation,
$\exp\left[-(\sum_i \lambda_i)^2/{2N}\right]=\sqrt{ {N\over {2\pi}}}\int_{-\infty}^{\infty} dz e^{iz 
\sum_i \lambda_i -Nz^2/2}$, to finally write the expected value of the product as,
\begin{equation}
{\left\langle \prod_{i}e^{i\lambda_i \psi_i}\right\rangle}_{\psi}=\sqrt{{N\over 
{2\pi}}}\int_{-\infty}^{\infty}dz e^{-Nz^2/2 + iz\sum_i \lambda_i -\sum_i \lambda_i^2/2}.
\label{HS}
\end{equation}
We then substitute Eq. (\ref{HS}) in Eq. (\ref{cosrep}) and carry out the Gaussian integrations over
the variables $\l_i$'s and $\lambda_i$'s to get
\begin{equation}
P_0(m,N)= \sqrt{{N\over {2\pi}}}\int_{-\infty}^{\infty} {{d\mu}\over {2\pi}}e^{i\mu m 
N}\int_{-\infty}^{\infty} dz e^{-Nz^2/2} {\left[A(z,\mu)\right]}^N,
\label{final0}
\end{equation}
where the function $A(z,\mu)$ is given by
\begin{equation}
A(z,\mu)={1\over {2}}\left[e^{i\mu} {\rm erfc}\left({z\over {\sqrt{2}}}\right)+e^{-i\mu}{\rm 
erfc}\left(-{z\over {\sqrt{2}}}\right)\right],
\label{azmu}
\end{equation}
with ${\rm erfc}(x) = {2\over {\sqrt {\pi}}}\int_x^{\infty} e^{-u^2}du$ being the complementary
error function.

We next expand the right hand side of Eq. (\ref{azmu}) in a binomial series, substitute
the resulting series in Eq. (\ref{final0}) and carry out the integration with respect 
to $\mu$ to obtain,
\begin{eqnarray}
\lefteqn{P_0(m,N)= {1\over {2^N}}\sqrt{{N\over {2\pi}}}\int_{-\infty}^{\infty}dz e^{-Nz^2/2} 
{N\choose {{(1-m)N}\over {2}}} \mbox{$\times$} } \quad\nonumber \\
&&\mbox{$\times$}{\left[{\rm erfc(z/\sqrt{2}})\right]}^{(1-m)N/2}
{\left[{\rm erfc(-z/\sqrt{2}})\right]}^{(1+m)N/2}.
\label{binomial}
\end{eqnarray}  
Keeping $m$ fixed we then use the Stirling's formula to approximate the combinatorial
factor in Eq. (\ref{binomial}) for large $N$ and then use the steepest descent method
to evaluate the integral in Eq. (\ref{binomial}) for large $N$. This gives, ignoring
pre-exponential factors, a similar asymptotic behavior as in the $1$-d case,
$P_0(m,N) \sim e^{-N \Theta(m)}$ where the large deviation function $\Theta(m)$ in this case 
is given by, 
\begin{equation}
\Theta(m)={1\over {2}}\min_z \left[ z^2 + \sum_{\sigma=-1,1}(1-m\sigma)\log\left(
{ {(1-m\sigma)}\over {2 {\rm erfc}\left({z\sigma}/\sqrt{2}\right)}}\right) \right].
\label{thetam}
\end{equation}   

In terms of the original variable $r=(1+m)/2$, the distribution is then given by
$P_0(r,N)= P_0(m=2r-1,N)/2\sim \exp\left[-N\Phi(r)\right]$ 
with $\Phi(r)=\Theta(2r-1)$ where
$\Theta(x)$ is given exactly by Eq. (\ref{thetam}). The function
$\Phi(r)$ is symmetric around $r=1/2$ since $\Theta(m)$ in Eq. (\ref{thetam})
is symmetric around $m=0$. As in the $1$-d case, it seems
difficult to obtain a closed form expression of $\Phi(r)$. However, it can be easily
evaluated from Eq. (\ref{thetam}) using Mathematica as displayed by the dotted line
in Fig.1 . 
Moreover, similar to the $1$-d case, one can evaluate $\Phi(r)$ analytically
near $r=1/2$ as well as near the tail regions $r\to 0,1$. Omitting the details
of algebra, we find,
\begin{eqnarray}
\Phi(r)&=& a + r\log\left({{br}\over {e}}\right) + \dots, r\to 0 \nonumber \\
&=& {{2\pi}\over {\pi+2}}{\left(r-{1\over {2}}\right)}^2 +\dots, r\to 1/2,
\label{limitingsk}
\end{eqnarray} 
where $a=\log (2) + z_0^2/2 -\log\left[{\rm erfc}\left(z_0/\sqrt{2}\right)\right]$,
$b= {\rm erfc}\left(z_0/\sqrt{2}\right)/ {{\rm erfc}\left(-z_0/\sqrt{2}\right)}$
and $z_0$ is the root of the equation,
\begin{equation}
z_0 + {\sqrt {2\over {\pi}}} { {e^{-z_0^2/2}}\over { {\rm erfc}\left(z_0/\sqrt{2}\right)}}=0.
\label{root1}
\end{equation}
Solving Eq. (\ref{root1}) numerically yields $z_0=-0.506054\dots$ which gives
$a=0.493919\dots$ and $b=2.26361\dots$. 

Thus the limiting behaviors of $\Phi(r)$ near $r=1/2$ and $r\to 0,1$ in the SK model
in Eq. (\ref{limitingsk}) are qualitatively similar to those in the $1$-d case
in Eq. (\ref{limiting1}). Using similar arguments as in Sec. I, one can understand
the behavior near $r=1/2$ as a consequence of a central limit theorem which predicts
a Gaussian behavior for $P_0(R,N)\sim \exp[-(R-N/2)^2/{2\sigma_N^2}]$, where $\sigma_N^2=\langle 
(R-N/2)^2\rangle$ is the variance. Comparing with Eq. (\ref{limitingsk}) we find
that the variance for large $N$ is given exactly by, 
$\sigma_N^2=(\pi+2)N/{4\pi}$. This result for the variance can also be derived 
by a direct method as shown in the Appendix, thus providing an
additional consistency check.
The results for the statistics of $R$ in the $1$-d case and in the SK model can
be jointly summarized as: The mean is always given by, $\langle R\rangle =N/2$ and the 
variance for large $N$ is given by,
\begin{eqnarray}
\sigma_N^2 &=& {5\over {12}}N, \quad\quad  \mbox{1-d} \nonumber \\
&=& { {\pi+2}\over {4\pi}}N, \quad \mbox{SK}.
\label{variance}
\end{eqnarray}.     

The behavior in the tail region near $r\to 0$ (or equivalently near $r\to 1$) is also
interesting. Let us, for simplicity, consider the $r\to 1$ limit where $R=N$.
It is clear from the expression of $P_0(N,N)$ in Eq. (\ref{gauge}) that 
the delta function will contribute only when each of the $\theta$ function
inside the sum are satisfied, i.e., if $\prod_{i} \theta\left(\sum_{j\ne i}\phi_{i,j}\right)=1$.
Thus evidently $P_0(N,N)={\left\langle \prod_{i} \theta\left(\sum_{j\ne 
i}\phi_{i,j}\right)\right\rangle}_{\phi}$. But this quantity is just the
average number of {\em metastable} configurations in the spin glass. To see
this clearly, consider again the spin glass Hamiltonian in Eq. (\ref{hamil}).
A spin configuration is called metastable if the energy required to flip
any of the $N$ spins is strictly positive. In other words, all the spins
must be aligned to their local fields in a metastable configuration.
Hence, for a fixed disorder, the fraction of metastable configurations (out of the total number 
of $2^N$ spin configurations) is given by,
$f(\vec J)= 2^{-N}\sum_{\vec S} \prod_{i} \theta(h_is_i)$ where $h_i$'s are the local fields.
Finally, the average (over disorder) fraction of the metastable configurations is given
by, ${\langle f(\vec J) \rangle}_{J}=\int f(\vec J)Q\left[\vec J\right] d\vec J$. 
Using once again the gauge transformation for Ising spins, $\phi_{i,j}=J_{i,j}S_iS_j$
one can easily express this average fraction as ${\langle f(\vec J) \rangle}_{J}={\left\langle 
\prod_{i} \theta\left(\sum_{j\ne
i}\phi_{i,j}\right)\right\rangle}_{\phi}=P_0(N,N)$. Our results on the large
deviation function near $r=0,1$ in Eq. (\ref{limitingsk}) indicates that
$P_0(N,N)=P_0(0,N) \sim e^{-a N}$ for large $N$ with $a=0.493919\dots$. 
On the other hand, the average number of metastable configurations in the SK 
model was computed long ago by Tanaka and Edwards\cite{TE} and also by Bray
and Moore\cite{BM} and this average is known to increase exponentially for large $N$
as $\sim e^{\alpha N}$ where $\alpha=0.1992$\cite{BM}. Hence the average fraction 
scales as $\sim e^{\alpha N}/2^N = e^{-c N}$ with $c=\log 2-\alpha=0.4919$.
Thus the constant $a$ in Eq. (\ref{limitingsk}) is precisely the same as
the constant $c$ and hence the limiting behavior of our large deviation function
near the tails $r=0,1$ is completely consistent with the calculation
of average number of metastable states.

Let us conclude this section with the following comment. In the case
of the $1$-d toy sequence, we found in Sec-II that the full
occupation time distribution $P(R,N)$ and consequently the
associated large deviation function is completely independent
of the distribution $\rho(\phi)$. In the case of the SK model,
we have derived the large devaition function for a specific
form of the disorder distribution, namely the Gaussian form.
Naturally the question arises as to how universal is this
large deviation function as one changes the disorder distribution.
Evidently for finite N the results in the SK case, unlike
the $1$-d case,  will depend
on the details of the distribution $\rho(J)$. 
However, due to the
$1/\sqrt{N}$ scaling in the definition of the  distribution of the
$J_{i,j}$, the large $N$ results inculding the large deviation  
are universal (upto rescaling by
a constant factor), provided the variance
of the $J_{i,j}$'s is finite. In the case of mean field spin glasses
with power law or L\'evy interactions\cite{CB}, the variance of the
$J_{i,j}$ is no longer finite and it would be interesting to study the
occupation time distribution in this context.    

\section{Summary and Conclusion}

The three main points of this paper are: (i) we have been able 
to derive, for the first time, 
exact results for the occupation time distribution of a non-Markovian process.
In our case, the stochastic process is not continuous in time, but rather a discrete
toy sequence. Nevertheless this toy sequence retains the non-Markovian property
which makes the results nontrivial. Besides the fact that exact results are always
useful and instructive, this toy 
sequence also appears
in various physical contexts such as diffusion process and spin glasses, thus extending the range 
of applications of our results. (ii) We also established an exact mapping of this 
sequence to an Ising spin glass chain using a gauge transformation. The 
occupation time distribution in the sequence then translates, via this mapping, into the 
distribution
of the thermally averaged number of spins that are aligned to their local fields
in the spin glass chain at infinite temperature when all spin configurations
are equally likely. This observation raises an interesting new question for 
any generic finite sized spin glass model: at a given temperature, what is the distribution of the 
thermally averaged 
number of locally aligned spins? Our exact
results in one dimension show that this distribution remains nontrivial even
at infinite temperature. (iii) We then were able to compute analytically this infinite temperature
distribution in the SK model of spin glasses with Gaussian disorder
and argued that for very large $N$, the associated large deviation function
is again universal, i.e., independent of the precise form of the
disorder distribution. 

We leave open the possibility of computing this 
distribution at a finite temperature for any spin glass model.
For example, it would be interesting to know how this distribution changes as one goes
below the spin glass transition temperature.    

The study of the number of metastable spins in various other spin glass models
is an open question. We mention a few cases where exact results along these
lines may be possible as the  average number of metastable
states is calculable: the SK model in the presence of external 
fields\cite{RODE}, p-spin spin glass models\cite{OF}, spin glasses on 
random graphs\cite{DELE}, mean field spin glasses with L\'evy interactions\cite{CB}, 
the Hopfield neural network model \cite{hop} and the Random 
Orthogonal Model\cite{ROM}. The study of spin glass models on random graphs 
of fixed connectivity $c$ are of particular interest as they interpolate between the 
one dimensional toy model studied here, at $c=2$, and the SK model in the 
limit $c\to \infty$.
   
\appendix
\section{Direct Calculation of the Variance of $M$}
In this appendix we compute the variances of the occupation time
both in the one dimensional toy sequence and in the SK model
by a more direct method. These results are identical
to those obtained from the limiting forms of the large deviation 
functions near $r=1/2$. 

We have in general,
\begin{eqnarray}
\sigma_N^2 &=& \langle \left[\sum_{i} {\left(\theta\left(\psi_i\right) - {1\over 2}\right)}^2
\right] \rangle \nonumber
\\ &=& {N\over 4} + 
2 \sum_{i<j} \left( \langle \theta(\psi_i) \theta(\psi_j) \rangle 
- {1\over 4} \right) \nonumber
\\ &=&  {N\over 4} + 
{1\over 2} \sum_{i<j} \langle {\rm sgn}(\psi_i)  {\rm sgn}(\psi_j)
\end{eqnarray}
 where we have made use of the identities $\langle \theta(\psi_i)\rangle =
1/2$ for all $i$ and ${\rm sgn}(x)  = 2\theta(x) -1$. 

In the one dimensional model only neighboring sites are correlated and
hence one has
\begin{eqnarray}
\sigma_N^2 &=& {N\over 4} + 
2 \sum_{i} \left( \langle \theta(\psi_i) \theta(\psi_{i+1}) \rangle 
- {1\over 4 }\right) \nonumber \\
&=& -{N\over 4} + 2N \langle \theta(\psi_1) \theta(\psi_{2})\rangle
\end{eqnarray}
where we have used the isotropy of the sites in the large $N$ limit.
One now has that
\begin{eqnarray}
& &\langle \theta(\psi_1) \theta(\psi_{2})\rangle = \nonumber \\ 
& &\int d\phi_0 d\phi_1 d\phi_2 \rho(\phi_0)\rho(\phi_1)\rho(\phi_2)
\theta(\phi_0 + \phi_1) \theta(\phi_1 + \phi_2) \nonumber \\
&=& \int_{-\infty}^\infty d\phi_0 \rho(\phi_0) \int_{-\phi_0}^\infty
d\phi_1 \rho(\phi_1) \int_{-\phi_1}^{\infty} d\phi_2 \rho(\phi_2)
\end{eqnarray}
We introduce the function $F(\phi) = \int_{-\phi}^{\infty} d\phi'\rho(\phi')$
and use the relations $\rho(\phi) = \rho(-\phi)$ and $dF/d\phi = \rho(\phi)$
to carry out the integration and thus obtain
\begin{equation}
\langle \theta(\psi_1) \theta(\psi_{2})\rangle = {1\over 3}
\end{equation}
Putting this altogether gives the large $N$ asymptotic result 
$\sigma_N^2 = 5N/12$. In fact using the generating function technique used
in this paper, and hence taking into account the boundary terms exactly,  
one can show that $\sigma_N^2 = 5N/12 -1/6$ for any $N$. Note that
this result, for arbitrary $N$, is
independent of the precise form of the distribution $\rho(\phi)$.

We now turn to the SK model where
$\psi_i = \sum_{j\neq i} \phi_{i,j}$ and the random variables 
$\phi_{i,j}$'s are independent and identically distributed with
the Gaussian distribution $\rho(\phi)=\sqrt{N/{2\pi}}e^{-N\phi^2/2}$.
Clearly, one has $\langle \psi_i^2\rangle = (1-1/N)$ and
$\langle \psi_i \psi_j\rangle = 1/N$. We next use the well known identity 
that holds only for 
Gaussian random variables,
\begin{equation} 
\langle {\rm sgn}(X) {\rm sgn}(Y) \rangle = {2\over \pi} \sin^{-1}\left( 
{\langle XY\rangle \over \sqrt{\langle X^2\rangle \langle Y^2\rangle}}\right).
\end{equation}
Using this identity we get for $i\neq j$,
\begin{equation}
 \langle {\rm sgn}(\psi_i) {\rm sgn}(\psi_j) 
\rangle = {2\over \pi} \sin^{-1}\left( {1\over N-1} \right).
\end{equation}
This yields
\begin{equation} 
\sigma_N^2 = {N\over 4} + {1\over 2\pi } N(N-1) 
\sin^{-1}\left( {1\over N-1} \right) \label{skvar}
\end{equation}
which gives the result $\sigma_N^2 = N(\pi + 2)/4\pi$ in the limit of 
large $N$. We note that the result in Eq. (\ref{skvar}) for finite $N$ is valid
only when the distribution of the $\phi_{i,j}$ is Gaussian. The finite $N$ result 
for arbitrary distribution of the $\phi_{i,j}$ will depend in general on the 
details of the distribution and hence, in contrary to what happens in the 
one dimensional toy model, will not be universal. However, as
argued in Sec. IV, the large $N$ results including the result
for the variance, i.e.,  
$\sigma_N^2 = N(\pi + 2)/4\pi $ is universal as long as the
variance of the $\phi_{i,j}$ is finite.

\end{multicols}

\end{document}